\documentstyle[12pt]{article}
\input epsf

\begin{document}
\baselineskip=15pt
\newcommand{\x}{{\bf x}}
\newcommand{\y}{{\bf y}}
\newcommand{\z}{{\bf z}}
\newcommand{\bp}{{\bf p}}
\newcommand{\A}{{\bf A}}
\newcommand{\B}{{\bf B}}
\newcommand{\p}{\varphi}
\newcommand{\del}{\nabla}
\newcommand{\be}{\begin{equation}}
\newcommand{\ee}{\end{equation}}
\newcommand{\bq}{\begin{eqnarray}}
\newcommand{\eq}{\end{eqnarray}}
\newcommand{\ba}{\begin{eqnarray}}
\newcommand{\ea}{\end{eqnarray}}
\def\r{\nonumber\cr}
\def\hf{\textstyle{1\over2}}
\def\qr{\textstyle{1\over4}}
\def\Sc{Schr\"odinger\,}
\def\sc{Schr\"odinger\,}
\def\'{^\prime}
\def\>{\rangle}
\def\<{\langle}
\def\-{\rightarrow}
\def\dbd{\partial\over\partial}
\def\tr{{\rm tr}}
\def\hg{{\hat g}}
\def\ca{{\cal A}}
\def\pd{\partial}
\def\dl{\delta}
\def\hp{\hat\phi}

\begin{titlepage}
\vskip1in
\begin{center}
{\large AdS/CFT boundary conditions and multi-trace perturbations
for fermions.}
\end{center}
\vskip1in
\begin{center}
{\large David Nolland}

\vskip20pt

Department of Mathematical Sciences

University of Liverpool

Liverpool, L69 3BX, England

{\it nolland@liv.ac.uk}

\end{center}
\vskip1in
\begin{abstract}

\noindent Extending the results of a previous paper, we consider
boundary conditions for spinor fields and other fields of non-zero
spin in the AdS/CFT correspondence. We calculate the RG-flow
induced by double trace perturbations dual to bulk spinor fields.
For spinors there is a half-unit shift in the central charge in
running from the UV to the IR, in accordance with the c-theorem.

\end{abstract}

\end{titlepage}

To investigate the AdS/CFT correspondence \cite{review} beyond
leading order in the large-$N$ expansion, we have to consider
quantum loops in the bulk theory. In general this includes string
loops, but to leading order in the $\alpha'$ expansion sub-leading
order corrections are given by Supergravity loops. To perform such
loop calculations requires a knowledge of what boundary conditions
to impose on the bulk fields.

In a recent paper \cite{bcs}, we discussed the possible boundary
conditions on scalar fields. In particular we calculated the
one-loop Weyl anomaly for all possible boundary conditions. The
asymptotic behaviour of the bulk scalar field has two components:

\be \phi=\alpha(x)z^{d-\Delta}+\beta(x)z^{\Delta}+\ldots,
\label{asympt}\ee where $\Delta$ is the larger root of the
equation $\Delta(\Delta-d)=m^2$. From a Hamiltonian point of view,
after a change of variables to reproduce the usual inner-product
on scalar fields, we found that diagonalising $\alpha$ and $\beta$
correspond to Dirichlet and Neumann conditions, respectively. For
Dirichlet conditions (and generic mixed boundary conditions) the
central charge is as computed in \cite{us,us2}. For Neumann
conditions, however, there is a unit shift of in the central
charge, relative to the central charge for Dirichlet conditions.

One situation in which the choice of boundary conditions is
important is when we perform double-trace perturbations of the
gauge theory that correspond to tachyonic fields in the
Supergravity theory \cite{witten}. These break the conformal
symmetry of the boundary gauge theory, and drive a renormalisation
group flow. For tachyonic modes whose masses lie in an appropriate
range, the difference between the ultraviolet and infrared fixed
points corresponds to the choice of Dirichlet or Neumann boundary
conditions for the bulk field \cite{bcs}.

In this letter we will extend these results to fermions and other
fields of non-zero spin. For fermions we will show that running
from the UV to the IR causes a shift of $\hf$ in the central
charge (we will relate this to boundary conditions for the
fermions). This shift in in accordance with the c-theorem
\cite{cardy,warner}.

We will work in an $AdS_{d+1}$ metric whose boundary is a
d-dimensional Einstein metric $\hat g$ \cite{us2}. The metric is

\be ds^2 = G_{\mu\nu}\,dX^\mu\,dX^\nu=dr^2 + z^{-2}¥\, e^{\rho}
\hg_{ij}(x)\, dx^i dx^j \, ,\quad e^{\rho/2}= 1-C\,z^{2}\,,
\quad¥C={l^2 {\hat R} \over 4\,d(d-1)}\,,\label{ads1} \ee

where $\hat R$ is the Ricci tensor on the boundary. The Euclidean
action for a spin-1/2 fermion in this metric is

\be \int d^{d+1}x\sqrt G\bar\psi(\gamma^\mu D_\mu-m)\psi.
\label{fac}\ee

The spin-covariant derivative is defined via the funfbein

\be V^\alpha_0={1\over z}\delta^\alpha_0,\qquad V^\alpha_i={1\over
z}e^{\rho/2}\tilde V^\alpha_i, \ee

where $\tilde V^\alpha_i$ is the vierbein for the boundary metric.
Making the change of variables $\psi=z^{d/2}e^{-d\rho/4
}\tilde\psi$ causes the volume element in the path-integral to
become the usual flat-space one, and the kinetic term in the
action acquires the usual form. The action can be written
(discarding a non-essential boundary term)

\be \int d^{d+1}x \bar{\tilde\psi}
\left(\gamma^0\partial_0+ze^{-\rho/2}\gamma^i\tilde
D_i-m\right)\tilde\psi. \ee

The $D_i$ derivative is spin-covariant with respect to the
boundary metric. We impose the following boundary conditions on
$\tilde\psi$:

\be Q_+\tilde\psi(0,x)=u(x)=Q_+u(x),\qquad\tilde\psi^\dagger(0,x)
Q_-=u^\dagger(x)=u^\dagger(x)Q_-, \label{bc1}\ee

for some local projection operators $Q_\pm$. The remaining
projections are represented by functional differentiation. The
partition function takes the form

\be \Psi[u,u^\dagger]=\exp[f+u^\dagger\Lambda u], \label{fw}\ee
and the Schr\"odinger equation that it satisfies can be written

\be {\partial\over\partial r_0}\Psi=-\int d^dx\left(u^\dagger
Q_-+{\delta\over\delta u}Q_+\right)h\left(Q_+
u+Q_-{\delta\over\delta u^\dagger}\right)\Psi, \label{fa}\ee

where $h=\tau e^{-\rho/2}\gamma^0\gamma^i\tilde D_i-\gamma^0m$. If
we make the specific choice $Q_\pm=\hf(1\pm\gamma^0)$, we can
write (\ref{fa}) as

\be {\partial\over\partial
r_0}\Psi=-\left[mu^\dagger{\delta\over\delta
u^\dagger}-m{\delta\over\delta u}u-\tau
e^{-\rho/2}u^\dagger\gamma\cdot\tilde Du+\tau
e^{-\rho/2}{\delta\over\delta u}\gamma\cdot\tilde
D{\delta\over\delta u^\dagger}\right]\Psi. \ee

Acting on (\ref{fw}) this implies that

\be \dot\Lambda=-2m\Lambda+\tau e^{-\rho/2}\gamma\cdot\tilde
D-\Lambda^2\tau e^{-\rho/2}\gamma\cdot\tilde D, \qquad\dot
f=\hf{\rm Tr}(-m+\Lambda\tau e^{-\rho/2}\gamma\cdot\tilde
D).\label{feq}\ee

Solving this in terms of Bessel functions we find that in momentum
space(assuming $m\ge0$)

\be \Lambda={I_{m+1/2}(p\tau)\over I_{m-1/2}(p\tau)}{\cal P},
\label{bf}\ee

where $\hf(1\pm{\cal P})$ are projectors onto +ve/-ve eigenvalues
of the operator $\gamma\cdot \tilde D$. As with scalar fields, to
get the correct scaling dimension as $\tau\to0$ requires
discarding terms of order less than $\tau^{2m}$ in the asymptotic
expansion of $\Lambda$\footnote{That this prescription is the
correct one is more easily seen if we construct the
wave-functional by integrating from $z=0$ to $z=\tau'$ with
$\tau'$ a large regulator, as in \cite{us}, giving the correct
scaling dimension as $\tau'\to\infty$ without any additional
renormalisation. The wave-functional tends to a delta functional
as $\tau'\to0$, as it should. The renormalisation we use here was
also discussed in \cite{volovich}.}. After removing the unwanted
terms, we have the asymptotic behaviour as $\tau\to0$

\be \Lambda\sim\tau^{2m}p^{2m}{\cal P}, \ee

and to get a finite wave-functional we must perform a
wave-function renormalisation $\psi\to\tau^{-m}\psi$. The scaling
dimension of the boundary field is then $d/2+m$ as required. We
could have achieved an identical result by diagonalising
$\tau^{m-d/2}Q_+\psi$ in the first instance; our approach has the
advantage of removing cutoff dependence from the functional
inner-product. Note that the choice $Q_\pm=\hf(1\pm\gamma^0)$ was
forced on us, and to consider other boundary conditions requires a
functional integration over boundary values. For example, if we
add a boundary term $\int (u^\dagger v-v^\dagger u)$ and integrate
over $u$ and $u^\dagger$ this changes the boundary conditions to

\be Q_-\tilde\psi(0,x)=v(x)=Q_-v(x),\qquad\tilde\psi^\dagger(0,x)
Q_+=v^\dagger(x)=v^\dagger(x)Q_+. \label{bc2}\ee

The renormalised fields $v$ and $v^\dagger$ represent the
canonical conjugates of the fields we diagonalised before. They
have scaling dimension $d/2-m$ and unitarity in the bulk indicates
that the boundary conditions (\ref{bc2}) are normalisable for
$m\le\hf$. For $m>1/2$ there is only one admissible boundary
condition. Recall that this assumed $m\ge0$.

A double-trace perturbation of the boundary theory corresponds to
adding a quadratic boundary term to the action:

\be I_{CFT}\to I_{CFT}+f\int u^\dagger u.\ee

 For spin-1/2
operators of scaling dimension $d/2+m$ where $|m|\le\hf$ this
drives a RG-flow from a UV fixed point at $f=0$ to an IR fixed
point at $f=\infty$. From the point of view of the holographically
dual bulk fields the UV fixed point corresponds to the boundary
conditions (\ref{bc1}) and the IR fixed point corresponds to the
boundary conditions (\ref{bc2}). A double-trace perturbation of
this kind was considered in \cite{petkou}.

To calculate the central charge associated with the Weyl anomaly
we expand (\ref{bf}) in terms of the positive-definite operator
$(\gamma\cdot\tilde D)^2$:

\be \Gamma=\gamma\cdot\tilde D\sum_{n=0}^\infty
d_n(r_0)(\gamma\cdot\tilde D)^{2n}. \ee

Notice that the coefficients $d_n$ {\em all} vanish as
$r_0\--\infty$. The equation (\ref{feq}) is easily solved in terms
of Bessel functions, but to regulate the expression for the free
energy $f$ we again use a heat-kernel expansion

\be {\rm Tr}(-m+\Gamma\tau\gamma\cdot\tilde
D)=\left(\sum_{n=0}^\infty d_n(r_0)\left(-{\partial\over\partial
s}\right)^{n+1}-m\right){\rm Tr}\exp\left(-s(\gamma\cdot\tilde
D)^2\right), \ee

where the heat-kernel has a Seeley-de Witt expansion

\be {\rm Tr}\,\exp \left(-s\left(-s(\gamma\cdot\tilde
D)^2\right)\right) =\int d^d x\,{\sqrt\hg}{1\over (4\pi s)^{d/2}
}\left(a_0+s\,a_1(x)
+s^2\,a_2(x)+s^3\,a_3(x)+..\right),\label{sdw} \ee
 The
contribution proportional to the $a_2$ coefficient is finite as
$s\-0$ and $r_0\--\infty$ and determines the anomaly, which is
therefore proportional to $m$. But since $m=\Delta-2$ we have as
before

\be \delta{\cal A}=-{\Delta-2\over 32\pi^2}\,a_2\,. \label{ar}\ee

This calculation of the anomaly assumed boundary conditions that
diagonalised $\tilde\psi^\dagger(0,x) \hf(1-\gamma^0)$ and
$\hf(1+\gamma^0)\tilde\psi(0,x)$, but we can change to other
boundary conditions by performing a functional integration over
the boundary values. For example to diagonalise
$\tilde\psi^\dagger(0,x) \hf(1+\gamma^0)$ and
$\hf(1-\gamma^0)\tilde\psi(0,x)$ we integrate over $u$ and
$u^\dagger$. There is a correction to the free energy given by

\be e^{2\delta f}=\det(\hf(1+\gamma^0)\Lambda). \ee

From this we calculate that

\be {\partial\over\partial r_0}\delta f={\rm
Tr}\left(\hf(1+\gamma^0){-2m\Lambda+\tau
e^{-\rho/2}\gamma\cdot\tilde D-\Lambda^2\tau
e^{-\rho/2}\gamma\cdot\tilde D\over2\Lambda}\right), \ee and
inserting the asymptotic behaviour of $\Lambda$ as $\tau\to0$ we
find the correction

\be {\partial\over\partial r_0}\delta f={\rm Tr}(-1/2),\ee leading
to the correction \be \delta{\cal A}=-{1/2\over 32\pi^2}\,a_2\,.
\ee

We conclude that the central charge is decreased by $1/2$ as we
run from the UV to the IR. Here we have used the definition of the
central charge given in \cite{cardy}, and the result is in
accordance with the proposed c-theorem in even dimensions
\cite{cardy, warner}.

We can also consider more general boundary conditions. If instead
of (\ref{bc2}) we impose the conditions

\be P_-\tilde\psi(0,x)=v(x)=P_-v(x),\qquad\tilde\psi^\dagger(0,x)
P_+=v^\dagger(x)=v^\dagger(x)P_+, \label{bc3}\ee for some local
projectors $P_\pm$, the correction to the free energy is given by

\be e^{2\delta f}=\det(\hf(1+\gamma^0)\Lambda+A^{-1}C), \ee with
$A=\{P_-,Q_+\}$ and $C=[P_-,Q_+]$. This leads to the same non-zero
correction to the central charge as before, if and only if $A$ or
$C$ vanishes. Otherwise there is no correction to the result
(\ref{ar}).

For spin-3/2 fermions the action can be written in the form
(\ref{fac}) (with a vector index on the fermion fields) and the
Schr\"odinger equation also takes the form (\ref{fa}). However,
unitarity considerations rule out the boundary conditions
(\ref{bc2}) for any value of the mass (in fact they rule out
$|m|<1/2$ completely).

For a scalar field the corresponding calculation was performed in
\cite{bcs} and we found a unit correction to the central charge.
Since bosonic fields of higher spin satisfy a Schr\"odinger
equation that has exactly the same form, there is a correction of
exactly the same amount, although unitarity severely restricts the
scaling dimensions for which a non-trivial flow is possible. So,
for example, in four dimensions there could in principle be a flow
for bulk vectors with masses saturating the BF-bound (ie.
$m^2=-1$) though such fields do not appear in the spectra of any
known supergravity compactifications.

It is interesting that choosing the "irregular" boundary
conditions causes the Weyl anomaly to vanish for the scalar and
spinor fields of the doubleton representation of SU(2,2/4). These
fields decouple from the spectrum and were not included in the
calculation of \cite{us2}. However they correspond to the
decoupled $U(1)$ factor that makes the boundary gauge group
$SU(N)$ instead of $U(N)$, and their contributions to the anomaly
should therefore be included if we think about interpolating
between large $N$ and $N=1$. This also happens for the doubleton
representation of OSp(8/4).

\end{document}